\documentclass[reprint,amsmath,aps,pra,footinbib,notitlepage,superscriptaddress]{revtex4-2}
\usepackage{graphicx,epstopdf,dcolumn,bm,mathtools,siunitx,multirow,hyperref,xcolor,amsfonts,physics,bibunits}
\usepackage[english]{babel}

\begin{document}


\title{Mobility Crossover in Two-Dimensional Berry Crystals}

\author{Zixuan Chai}
\affiliation{%
 Department of Physics, University of Cambridge, Cambridge, UK
}%
\affiliation{%
 Department of Physics, Harvard University, Cambridge, Massachusetts 02138, USA
}%

\author{Si-Yuan Chen}
\affiliation{%
 Department of Physics, Harvard University, Cambridge, Massachusetts 02138, USA
}%
\affiliation{%
 International Center for Quantum Materials and School of Physics, Peking University, Beijing 100871, China
}%

\author{Chenzheng Yu}
\affiliation{%
 International Center for Quantum Materials and School of Physics, Peking University, Beijing 100871, China
}%
\affiliation{%
 Department of Chemistry and Chemical Biology, Harvard University, Cambridge, Massachusetts 02138, USA
}%

\author{Anton M.~Graf}
\affiliation{Harvard John A. Paulson School of Engineering and Applied Sciences,
Harvard, Cambridge, Massachusetts 02138, USA}
\affiliation{%
 Department of Chemistry and Chemical Biology, Harvard University, Cambridge, Massachusetts 02138, USA
}%

\author{Eric J. Heller}
\affiliation{%
 Department of Chemistry and Chemical Biology, Harvard University, Cambridge, Massachusetts 02138, USA
}%

\affiliation{%
 Department of Physics, Harvard University, Cambridge, Massachusetts 02138, USA
}%

\author{Joonas~Keski-Rahkonen}
\email{joonas_keskirahkonen@fas.harvard.edu}
\affiliation{%
 Department of Chemistry and Chemical Biology, Harvard University, Cambridge, Massachusetts 02138, USA
}%
\affiliation{%
 Department of Physics, Harvard University, Cambridge, Massachusetts 02138, USA
}%

\begin{abstract}

Anderson localization is commonly regarded as a generic fate of wave propagation in disordered media, yet underlying spatial correlations can profoundly reshape this picture. Here we introduce a class of two-dimensional correlated potentials generated by random superpositions of plane waves with fixed wavevector magnitude. Despite their simplicity, these \emph{Berry crystals} arise across a wide range of physical settings and realize a distinct, largely unexplored regime between uncorrelated disorder and quasicrystalline order. In stark contrast to the standard Anderson scenario, this localization transition is not governed by a sharp mobility edge but by an extended regime of critical, multifractal states that mediates an inverted passage from extended to localized behavior. We develop a quantitative framework for this mobility crossover by mapping the system onto an effective Anderson model on a Bethe lattice. More broadly, our results establish Berry crystals as a versatile platform for exploring correlation-driven mobility crossovers and the impact of spatial structure of disorder on Anderson localization.

\end{abstract}

\maketitle

\noindent
\paragraph*{Introduction.---}
The physics of Anderson transitions between localized and metallic phases remains a central interest across diverse systems~\cite{Anderson_localization_book, Evers_Rev.Mod.Phys_80_1355_2008}. In the standard uncorrelated-disorder narrative~\cite{Anderson_Phys.Rev_109_1492_1958, Abrahams_Phys.Rev.Lett_42_673_1979}, diffusion is fully suppressed in one and two dimensions, with a metal–insulator transition occurring only in three dimensions. However, the subtle quantum interference driving the localization can be readily disrupted. For instance, many-body interactions typically destabilize Anderson localization~\cite{Basko_Ann.Phys_321_1126_2006, Gornyi_Phys.Rev.Lett_95_206603_2005}, while simultaneously enabling their own distinct form of localization~\cite{Abanin_Rev.Mod.Phys_91_021001_2019}. Conversely, slowly morphing disorder can lead to only evanescent, cyclic buildup of Anderson localization (for capturing the essential aspects of this phenomenon, see, e.g., Refs.~\cite{Di.Sante_Phys.Rev.Lett_118_036602_2017, Pustogow_nat.commun_12_1571_2021, Fratini_Adv.Funct.Mater_26_2292_2016}), allowing diffusion to progress but at the Planckian rate~\cite{Broken_Anderson_localization}, as particularly suggested for thermal lattice vibrations~\cite{Donghwan_phys.rev.b_106_054311_2022, zimmermann_entropy_26_552_2024} and subsequently linked to strange-metal behavior~\cite{aydin_pnas_121_e2404853121_2024, KeskiRahkonen_phys.rev.lett_132_186303_2024, Fratini_Phys.Rev.Lett_136_086301_2026}. In addition to these two scenarios, spatial correlations can also fundamentally alter the Anderson paradigm (see, e.g., Refs.~\cite{Izrailev_Phys.Rev.Lett_82_4062_1999, de.Moura_Phys.Rev.Lett_81_3735_1998, Dunlap_Phys.Rev.Lett_65_88_1990, Marcantoni_Phys.Rev.B_106_134211_2022, Titov_Phys.Rev.Lett_95_126602_2005, Fratini_Phys.Rev.A_92_063621_2015, Mogilevtsev_Phys.Rev.B_84_094204_2011, Semeghinit_Nat.Phys.11_554_2015, Stephens_Chem.Phys_177_727_1993, Russ_Physica.A_266_492_1999, Schwartz_Nature_446_52_2007}), such as by generating mobility edges that host multifractal critical states which are extended yet non-ergodic~\cite{Wegner_Z.Phys.B_36_209_1980, Obuse_Phys.Rev.B_82_035309_2010, Mirlin_Phys.Rev.B_62_7920_2000, Mirlin_Phys.Rep_326_259_2000,Castellani_J.Phys.A:Math.Gen_19_429_1986}.

Quasicrystals provide an emblematic solid-state example of such correlation-driven complications (see, e.g., Refs.~\cite{Sokoloff_Phys.Rep_126_189_1985, Poon_Adv.Phys_41_303_1992}): The influence of long-range order combined with lack of translational symmetry gives rise to intriguing localization properties in quasiperiodic systems. The existence of mobility edges related to Anderson localization has been experimentally confirmed in both real quasicrystals~\cite{Sarkar_Phys.Rev.B_103_L241106_2021} as well as engineered platforms, such as photonic analogs~\cite{Jeon_Nat.Phys_13_363_2017, Levi_Science_332_6037_2011, Lahini_Phys.Rev.Lett_103_013901_2009, Kohmoto_Phys.Rev.Lett_58_2436_1987}. At the principal level, the Aubry–André model~\cite{Aubry_Ann.Israel.Phys.Soc_3_18_1980} demonstrates a self-dual transition between fully extended and localized phases without a mobility edge, but breaking this duality generically induces mobility edges and coexistence of localized and extended states~\cite{Biddle_Phys.Rev.Lett_104_070601_2010, Deng_Phys.Rev.Lett_123_025301_2019, Ganeshan_Phys.Rev.Lett_114_146601_2015, Wang_Phys.Rev.Lett_125_196604_2020, Das.Sarma_Phys.Rev.Lett_61_2144_1988, Biddle_Phys.Rev.A_80_021603_2009, Li_Phys.Rev.B_96_085119_2017, Roy_Phys.Rev.Lett_126_106803_2021}. More broadly, the commonly intertwined presence of critical states~\cite{Yao_Phys.Rev.Lett_123_070405_2019, Roy_Phys.Rev.B_103_075124_2021, Goncalves_Phys.Rev.Lett_131_186303_2023, Fraxanet_Phys.Rev.B_106_024204_2022, Wang_Phys.Rev.B_106_L140203_2022, Duncan_Phys.Rev.B_109_014210_2024, Goblot_Nat.Phys_16_832_2020, Tong_SciPost.Phys_12_027_2022} and mobility edges~\cite{Sbroscia_Phys.Rev.Lett_125_200604_2020,Chan_Phys.Rev.Lett_80_956_1998, Levine_Phys.Rev.Lett_53_2477_1984, Zhu_Phys.Rev.A_109_013314_2024, Wang_Nat.Photon_18_224_2024} are reported across a wide variety of quasicrystals and generic quasiperiodic potentials, with physical realizations ranging from photonic crystal~\cite{Lahini_Phys.Rev.Lett_103_013901_2009} to superconducting qubits~\cite{Li_npj.Quantum.Inf_9_40_2023, Roushan_Science_358_1175_2017}. These results underscore the conclusion that critical states linked to mobility edges seem to emerge as a robust feature of correlated systems, underlining a conclusion that spatial structure, not randomness alone, governs localization.

In this Letter, we introduce a qualitatively distinct circumstance residing within the unexplored intermediate terrain of amorphous random disorder and quasicrystal order: a two-dimensional aperiodic potential with long-range correlations lacking quasiperiodicity constructed as a random superposition of plane waves with fixed wavevector magnitude. Such potentials arise naturally in wave-chaotic cavities (see, e.g., Refs~\cite{Stein_Phys.Rev.Lett_68_2867_1992, Kudrolli_Phys.Rev.Lett_75_822_1995, Stockmann_book}), and bear close parallel to powder diffraction, where a Berry wave structure is shown to support Bragg scattering despite the absence of crystal order~\cite{Kim_Phys.Rev.Lett_128_200402_2022}. In the sphere of condensed matter, we term this structure a \emph{Berry crystal}, reflecting its origin in quantum chaos~\cite{Berry_J.Phys.A:Math.Gen_10_2083_1977, Longuet-Higgins_Philos.Trans.R.Soc.A_249_321_1957}. Moreover, kindred disorder but multichromatic potentials emerge commonly, for instance, in the circumstances of blackbody radiation as well as lattice vibrations, as originally presented by Hanbury Brown and Twiss~\cite{Hanbury_Proc.R.Soc.A_242_300_1957}, and by Bardeen and Shockley~\cite{Bardeen_Phys.Rev_80_72_1950, Shockley_Phys.Rev_77_407_1950}, respectively.

We demonstrate that a Berry crystal hosts two distinct types of mobility crossovers in contrast to a sharp mobility edge commonly expected: a single extended–critical–extended crossover that bypasses localization, and a dual Anderson transition between extended and localized states separated by a broad critical-state window. Furthermore, we show that our physical setting maps onto a Cayley-tree model further connected to an effective Bethe lattice, shedding light onto a microscopic mechanism for the emergence of this counterintuitive behavior, while offering a quantitative description of the crossover boundaries.

\paragraph*{Berry crystal.---}
Specifically, we consider the following single-particle Hamiltonian given in atomic units as
\begin{equation}\label{Eq:Hamiltonian}
    H = -\frac{\hbar^2}{2m}\nabla^2 + V(\mathbf{r};\{\phi_j\}),
\end{equation}
where the Berry potential defined as
\begin{align}
    V(\mathbf{r};\{\phi_j\})=\frac{\beta}{\sqrt{N/2 \;}}\; \sum_{j=1}^N\cos(\mathbf{q}_j\cdot\mathbf{r}+\phi_j), 
    \label{eq:randompot}
\end{align}
with the (effective) mass $m$, the potential strength $\beta$, and the number of modes $N$. Mimicking the concept behind the famous Berry conjecture~\cite{Berry_J.Phys.A:Math.Gen_10_2083_1977}, the wavevectors $\mathbf{q}_j = q(\cos\theta_j, \sin\theta_j)$ have fixed magnitude $q$, while the orientations $\theta_j$ and phases $\phi_j$ are independently and uniformly distributed over $[0,2\pi]$.

The potential defined above belongs to a special class known as statistically homogeneous disorder~\cite{Ziman_book}, meaning that its spatial (also ensemble) average vanishes, $\langle V(\mathbf{r})\rangle = 0$. Nevertheless, the potential correlations extend across all length scales, as indicated by its autocorrelation function
\begin{equation}
    \mathcal{C}(\delta\mathbf{r}) = \langle V(\mathbf{r}) V(\mathbf{r} + \delta \mathbf{r})\rangle = \frac{\beta^2}{N} \sum_{j = 1}^N \cos ( \mathbf{q}_j \cdot \delta \mathbf{r}),
\end{equation}
which depends only on the relative displacement $\delta \mathbf{r}$, reflecting translational invariance. The disorder strength itself is characterized by the root-mean-square amplitude $V_{\mathrm{rms}} = \sqrt{\mathcal{C}(0)} = \beta$. 

It is precisely this long-range correlated structure that sharply distinguishes the Berry potential from the standard Anderson setting~\cite{Anderson_localization_book, Evers_Rev.Mod.Phys_80_1355_2008}, which typically assumes either uncorrelated (white-noise) disorder, $\mathcal{C}(\delta \mathbf{r}) = 0$ for $\delta \mathbf{r} \neq 0$, or short-range correlations decaying exponentially, $\mathcal{C}(\delta \mathbf{r}) \propto e^{-|\delta \mathbf{r}|}$. At the same time, even though Berry crystals share a self-similar structure exhibiting long-range correlations without periodicity, they generally lack the discrete rotational symmetries characteristic to quasicrystalline order, such as the fivefold symmetry of Penrose tilings~\cite{Marder_book} or the tenfold symmetry observed in the first quasicrystals~\cite{Shechtman_Phys.Rev.Lett_53_1951_1984}. In the limit $N \to \infty$, the correlation function reduces to $\mathcal{C}(\delta \mathbf{r}) = \beta^2 J_0(q |\delta \mathbf{r}|)$, where $J_0$ is the zeroth-order Bessel function.~\cite{Kim_Phys.Rev.Lett_128_200402_2022} While short-range order persists, the correlations decay algebraically at large distances as $\mathcal{C}(\delta \mathbf{r}) \propto |\delta \mathbf{r}|^{-1/2}$ for $q|\delta \mathbf{r}| \to \infty$. Consequently, a Berry crystal situates within an unexplored territory bridging quasicrystal behavior and Anderson localization.

Experimentally, Berry-crystal-like potentials arise naturally in optical cavities with rough or ballistically chaotic boundaries, where single- or multimode lasing generates wave fields composed of random superpositions of plane waves propagating in different directions~\cite{Stein_Phys.Rev.Lett_68_2867_1992, Kudrolli_Phys.Rev.Lett_75_822_1995, Stockmann_book}. Related realizations are available in a variety of other platforms, including ultracold atoms in optical lattices, waveguide arrays, photonic lattices, and programmable quantum simulators, the latter of which have already been utilized to study many-body localization~\cite{Schwartz_Nature_446_52_2007, Billy_Nature_453_891_2008, Lahini_Phys.Rev.Lett_100_013906_2008, Roati_Nature_453_895_2008, Smith_Nat.Phys_12_907_2016, Schreiber_Science_349_842_2015}. These platforms offer substantial control over the strength and structure of the underlying potential. On the other hand, the Berry potential is closely related to the limiting case of fully overlapping crystallites with all possible orientations in powder x-ray diffraction~\cite{Kim_Phys.Rev.Lett_128_200402_2022}. Beyond such synthetic settings, including multilayer graphene systems~\cite{Ahn_Science_361_782_2018, Uri_Nature_620_762_2023}, it nevertheless remains an open question whether Berry-type structures can describe natural materials, such as amorphous solids. Finally, in Appx.~\ref{Appx:experimental_scheme}, we outline a complementary and experimentally accessible route for engineering Berry potentials employing cold atoms materials.  

From a theoretical standpoint, we investigate the Berry crystal by computing hundreds of eigenstates $\psi_n$ and eigenenergies $\varepsilon_n$ of the Hamiltonian in Eq.~\eqref{Eq:Hamiltonian} in the vicinity of a prescribed target energy $E$, as detailed in Appx.~\ref{Appx:spectrum}, while systematically varying the potential strength $\beta$ as well as the number of modes $N$. Unless otherwise stated, we take $q=0.5\pi \, \textrm{Å}^{-1}$ and $m$ to be the bare electron mass. Throughout this Letter, we express energies, such as $\beta$, in units of $E_0  = \hbar^2q^2/(8m)\approx2.35 \, \textrm{eV}$, set by the resonant scattering condition, and length scales, such as the system size $L$, in units of $\lambda = 2\pi/q=4\, \textrm{Å}$. However, we note that the physical conclusions presented below are not specific to these particular parameter choices.
 
\paragraph*{Eigenstate morphology.---}

We characterize the potential emergence of Anderson localization in our Berry crystal employing the generalized inverse participation ratio (IPR)~\cite{Girvin_book},
\begin{equation} \label{Eq:IPR_definition}
   \Gamma_q^{(n)} =  \int |\psi_n(\mathbf{r})|^{2q} d\mathbf{r} \propto L^{-D_q(q-1)}
\end{equation}
which then defines the generalized fractal dimension $D_q$ governing the scaling of the IPR with system size $L$. In the following, we focus on the second moment ($q = 2$) that provides a well-established diagnostic for distinguishing different spectral regimes (see, e.g., Ref.~\cite{Fartash_Phys.Rev.B_113_195401_2026}). Specifically, we determine $D_2(E)$ for $\beta\in[0.2,1.0]$ and $N\in[10,120]$ from the finite-size scaling of the ensemble-averaged IPR, following the procedure detailed in Appx.~\ref{Appx:scaling_analysis}. We first consider the representative case $N=60$ to expose the generic behavior and subsequently examine its dependence on $N$, guided by the theoretical framework developed below.

\begin{figure}[h!]
    \centering
    \includegraphics[width=0.45\textwidth]{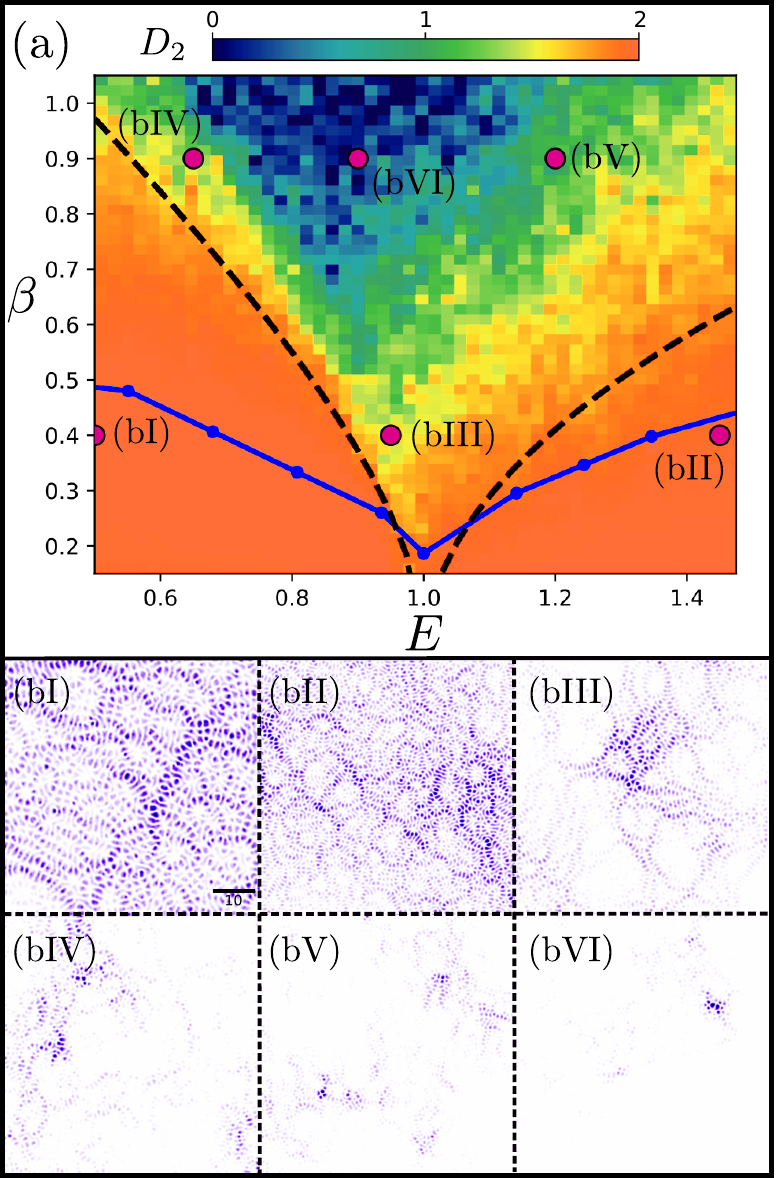}
    \caption{Mobility crossovers in a Berry crystal: (a) Fractal dimension $D_2$ [see Eq.~\eqref{Eq:IPR_definition}] from the inverse participation ratio analysis as a function of potential strength $\beta$ and energy $E$ for $N=60$, revealing two distinct crossover scenarios. At weak potential strengths ($\beta \le 0.5$), extended states undergo a transition mediated by a broad critical regime near resonant energy $E=1$, while at stronger potentials ($\beta \ge 0.7$), a dual crossover involving Anderson-localized states embedded within wide bands of critical states. Solid blue and black dashed curves indicate crossover boundaries estimated from perturbation theory and effective Bethe-lattice mapping, respectively. (bI)–(bVI) Representative eigenstates in the extended, critical, and localized regimes corresponding to the points marked in (a).}
    \label{Fig:phase_diagram}
\end{figure} 

Figure~\ref{Fig:phase_diagram}(a) presents $D_2$ as a function of $E$ and $\beta$ for $N = 60$, revealing three distinct regimes corresponding to extended, critical, and localized states. Extended (ergodic) states  follow the scaling $\Gamma_2 \propto L^{-2}$ and hence $D_2 \simeq 2$ [Figs.~\ref{Fig:phase_diagram}(bI)–(bII)]. At the opposite limit, Anderson-localized states with localization length $\xi$ become insensitive to the system size for $L \gg \xi$, yielding $D_2\simeq 0$ [Fig.~\ref{Fig:phase_diagram}(bVI)]. Between these two limits, we find critical states characterized by $0 < D_2 <2$ [Figs.~\ref{Fig:phase_diagram}(bIII)–(bV)]. These states are neither extended nor (Anderson) localized, but instead exhibit strong spatial fluctuations and scale-free structure, reflected in their nontrivial IPR scaling. Unlike the standard scenario of Anderson transition, where this type of criticality is typically restricted to a narrow mobility edge, the critical states of the Berry crystal span across a broad energy window, forming an extended critical regime between the extended and localized phases. 

This phase structure brings about two distinct types of mobility crossover seen in Fig.~\ref{Fig:phase_diagram}(a). For weak Berry potentials ($\beta\lesssim0.7$), low-energy extended states evolve into a broad critical regime around the resonant scattering condition $E\sim1$ and subsequently recover their extended character at higher energies, without developing an appreciable localized regime. The resulting crossover therefore connects two extended regimes through an intermediate band of critical states. As $\beta$ increases, this critical regime broadens until localized states first emerge at $\beta\approx0.75$. For $\beta\gtrsim0.8$, the crossover consequently acquires a dual structure: low-energy extended states evolve into localized states at intermediate energies and then return to an extended regime at higher energies, with the localized region separated from the extended regimes by broad bands of critical states.

The probability density profiles of representative eigenstates sampled along the two crossover trajectories marked in Fig.~\ref{Fig:phase_diagram}(a) offer a direct coordinate-space view of the three distinct regimes. The low- and high-energy states on either side of the crossovers are extended [Figs.~\ref{Fig:phase_diagram}(bI)–(bII)], spreading ergodically throughout the available space with the characteristic structure of chaotic eigenstates anticipated by the Berry conjecture~\cite{Berry_J.Phys.A:Math.Gen_10_2083_1977}. The intervening critical states [Figs.~\ref{Fig:phase_diagram}(bIII)–(bV)] instead display sparse, filamentary structures with pronounced fluctuations across a broad range of length scales, providing a coordinate-space manifestation of their multifractal character. Finally, in the dual-crossover regime at larger $\beta$, an island linked to localized states emerges in the vicinity of the resonant energy, flanked by critical regimes on either side. As illustrated in Fig.~\ref{Fig:phase_diagram}(bVI), these states are concentrated around a single dominant localization center, with the probability density decaying rapidly away from it and leaving most of the available space effectively unoccupied.

\paragraph*{Spectral statistics.---}
From the perspective of spectral correlations, we probe the mobility crossover considering the ratio of consecutive level spacings, which provides a compact, unfolding-free diagnostic of spectral statistics~\cite{Atas_Phys.Rev.Lett_110_084101_2013}. In the spirit of the Bohigas–Giannoni–Schmit conjecture~\cite{Bohigas_phys.rev.lett_52_1_1984}, the mean gap ratio distinguishes Poisson statistics, associated with quantum-regular spectra, from Wigner–Dyson statistics of the Gaussian orthogonal ensemble (GOE), expected for time-reversal-symmetric chaotic systems. We accordingly introduce a normalized chaoticity indicator $\chi$ based on the gap ratio, as detailed in Appx.~\ref{Appx:level_statistics}, with $\chi=0$ and $\chi=1$ corresponding to the Poisson and GOE limits, respectively.

\begin{figure}[h!]
    \centering
    \includegraphics[width=0.5\textwidth]{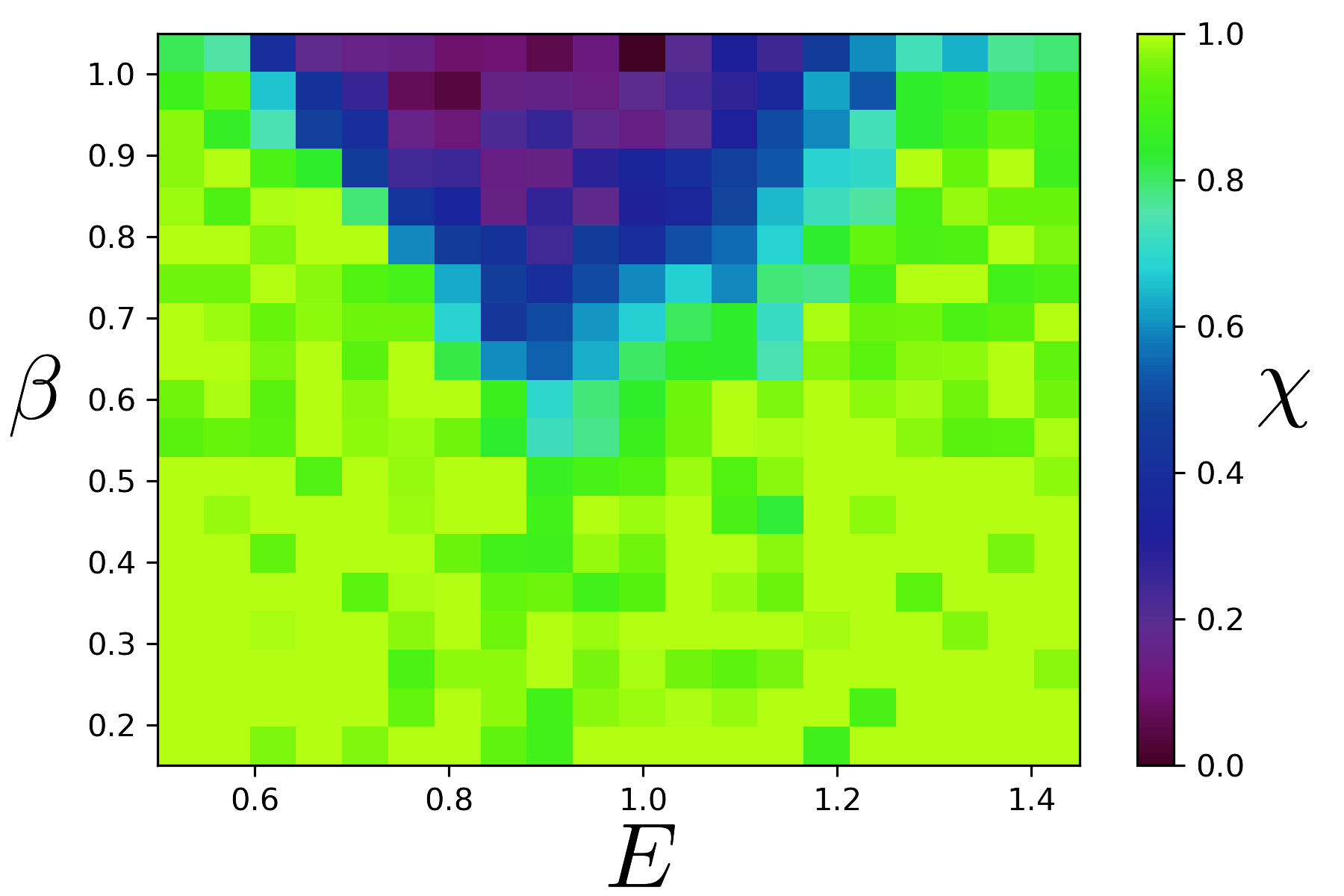}
    \caption{Spectral statistics and chaos indicator: chaos indicator $\chi$, defined via the gap ratio, as a function of potential strength $\beta$ and energy $E$. While $\chi$ detects the crossover between localized ($\chi \sim 0$) and extended ($\chi \sim 1$) states, it compresses this feature into a sharp, pseudo-Anderson transition above $\beta \ge 0.6$. By misclassifying most critical states as extended, the spectra-statistics analysis masks both the dual-transition structure and the low-$\beta$ crossover revealed by the IPR analysis.}
    \label{Fig:level_statistics}
\end{figure}

Level statistics has in turn been demonstrated to be able to distinguish Anderson-localized from ergodic extended states~\cite{Izrailev_Phys.Rep_196_299_1990}: localized states exhibit Poisson statistics ($\chi \sim 0$), whereas ergodic extended states display GOE statistics ($\chi \sim 1$). Critical states, however, need not be uniquely identified by this spectral diagnostic. Their nonergodic character can lead to intermediate spectral statistics, combining level repulsion at small spacings with a Poisson-like tail at large spacings. We find that this limitation is particularly relevant for the Berry crystal. Figure~\ref{Fig:level_statistics} shows the chaos estimator $\chi$ as a function of energy $E$ for several values of $\beta$, paralleling the parameter range of Fig.~\ref{Fig:phase_diagram}. While the measure $\chi$ captures the overall evolution across the mobility crossover, it primarily discerns extended from localized behavior, and assigns most critical states to the latter. Consequently, the crossover appears substantially sharper than in the IPR analysis of Fig.~\ref{Fig:phase_diagram}. In particular, the broad critical regime associated with the crossover at low $\beta \lesssim 0.5$ is completely absent in the context of the spectral statistics, whereas the dual crossover for $\beta \gtrsim 0.6$ appears deceptively similar to an ordinary Anderson transition with a mobility-edge boundary.

We obtain the same qualitative picture utilizing the spectral variance as an alternative measure~\cite{Ramos_Phys.Rev.C_95_054326_2017}, as detailed in Appx.~\ref{Appx:level_statistics}. The limitations of spectral statistics in resolving the critical regime are further highlighted by the wavepacket analysis presented in Appx.~\ref{Appx:Wavepacket_analysis}, which gives another probe of the mobility crossover. Whereas the level statistics in Fig.~\ref{Fig:level_statistics} suggest an ostensible Anderson transition, the wavepacket investigation supports the broader crossover scenario revealed by the IPR examination in Fig.~\ref{Fig:phase_diagram}, including the extended critical regimes intrinsic to the Berry crystal. 

\paragraph*{Crossover estimators.---}
We next estimate the critical potential strength $\beta$ marking the onset of the mobility crossover from the viewpoints of two complementary approaches: perturbation theory and a mapping to an effective Bethe lattice. Here we summarize the key ideas underlying these two approaches, with detailed derivations and analyses are provided in Appx.~\ref{Appx:Perturbation_theory} and Appx.~\ref{Appx:Bethe_lattice}.

Within the perturbative approach, we treat the Berry potential in Eq.~\eqref{eq:randompot} as a perturbation of the plane-wave states $\lvert\mathbf{k}\rangle$ and expand an eigenstate of the Hamiltonian in Eq.~\eqref{Eq:Hamiltonian} as $\lvert\psi\rangle\approx\sum_{\mathbf{k}}c_{\mathbf{k}}\lvert\mathbf{k}\rangle$. Convergence of the resulting expansion, i.e., $\sum_{\mathbf{k}}|c_{\mathbf{k}}|^2<\infty$ implies that the eigenstate remains continuously connected to the plane-wave basis and is therefore extended. Conversely, a breakdown of this expansion signals the loss of this perturbative connection, yielding an indication of spatial localization. 

We identify the boundary between the extended and critical regimes with the point at which the perturbative expansion first becomes numerically divergent, shown by the solid blue curve in Fig.~\ref{Fig:phase_diagram}(a). The expansion converges at both low and high energies but breaks down in the vicinity of the scattering resonance, $E \sim 1$. Although this criterion captures the overall location of the mobility crossover, its quantitative accuracy depends on energy: it underestimates the critical $\beta$ for $E\lesssim1$, while showing improved agreement for $E \gtrsim 1$. 

To refine the estimate of the mobility crossover, we construct an effective Bethe-lattice description of the Berry crystal. Since the Berry potential primarily induces momentum-space scattering through its characteristic wavevector $\mathbf q$, we can map the problem onto a tight-binding model on a highly connected graph of plane-wave states $\lvert \mathbf k \rangle$. For near-resonant processes $\mathbf k \rightarrow \mathbf k + \mathbf q$, the structure of Eq.~\eqref{eq:randompot} strongly suppresses closed loops, hence generating an approximately tree-like momentum-space connectivity. Retaining the dominant scattering processes in momentum space, the Hamiltonian in Eq.~\ref{Eq:Hamiltonian} maps onto an effective Cayley tree,
\begin{equation}
H_{\rm eff} = \sum_{i = 1}^N \tilde{\Delta}_{i} |i\rangle \langle i| + \sum_{\langle i,j\rangle} e^{i\phi_{ij}} |i\rangle\langle j| + \textrm{h.c},
\end{equation}
where $|i\rangle$ are the momentum space states spanning the graph, $\phi_{ij}$ are random phases, and $\tilde{\Delta}_{i}$ are onsite energies determined by the corresponding scattering processes depending on $E$ and $N$.

In practice, we model the onsite energies $\tilde{\Delta}_i$ as random variables drawn from a probability distribution with variance
\begin{equation}\label{eq:BL_critical}
W(E, \beta) \approx
\begin{cases}
\dfrac{46\pi^2}{N \beta^2} (2 - \sqrt{E})(1 - \sqrt{E}), & E < 1, \\
\\
\dfrac{16\pi}{5\beta}\sqrt{\dfrac{2(E - 1)}{N}}, & E > 1.
\end{cases}
\end{equation}
The resulting treatment of our effective model is therefore analogous to an Anderson model on a Bethe lattice with connectivity $K=3$. Furthermore, such models exhibit an Anderson transition at a critical disorder strength $W_C$, whose value depends on the variance of disorder, like established for uniform and Cauchy-type disorders~\cite{Altshuler_Phys.Rev.Lett_117_156601_2016, Kravtsov_Ann.Phys_389_148_2018}. Following these results, we take $W_C \sim 6$ for the present model, yielding the crossover boundary $W(E,\beta)=W_C$ shown by the black dashed curve in Fig.~\ref{Fig:phase_diagram}(a). This criterion substantially improves upon the perturbative estimate [blue curve in Fig.~\ref{Fig:phase_diagram}(a)] away from the resonance energy $E \sim 1$, while the perturbative approach remains more accurate in its immediate vicinity.

\begin{figure}[h!]
    \centering
    \includegraphics[width=0.45\textwidth]{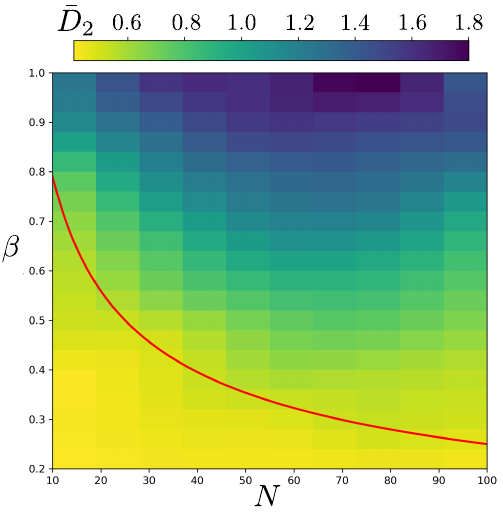}
    \caption{Mean fractal dimension $\bar{D}_2$ as a function of potential strength $\beta$ and mode number $N$, averaged over the energy window $E \in [0.5, 1.5]$ shown in Fig.~\ref{Fig:phase_diagram} (a). The red curve illustrates the analytical scaling prediction $\beta_c \propto N^{-1/2}$ derived from the effective Bethe-lattice model.}
    \label{Fig:phase_diagram_2}
\end{figure}

The Bethe-lattice mapping further yields a simple prediction for the dependence of the critical potential strength on the number of modes, namely $\beta_c\propto N^{-1/2}$. We test this scaling by evaluating the energy-averaged fractal dimension $\bar D_2(\beta,N)$ over $\Delta E=[0.5,1.5]$, the same energy window used in Fig.~\ref{Fig:phase_diagram}(a). As shown in Fig.~\ref{Fig:phase_diagram_2}, $\bar D_2$ decreases with increasing $\beta$, signaling a crossover from extended to localized states. In agreement with the proposed scaling derived from Eq.~\eqref{eq:BL_critical}, the crossover shifts to progressively smaller $\beta$ as $N$ increases, with the predicted scaling shown by the red line. The scaling remains well described up to $N=60$ (the representative case in Fig.~\ref{Fig:phase_diagram_2}), but begins to deviate for $N \gtrsim 100$. This deviation is expected as higher-order scattering processes become increasingly important, heralding the failure of our effective Bethe-lattice description for large $N$.   

\paragraph*{Conclusion.---}
In general, our results reveal a complexity in Berry crystals that is not apparent from the deceptively simple form of Eq.~\eqref{Eq:Hamiltonian}, placing them beyond conventional paradigms of disorder. Because the underlying physics arises from linear wave propagation in correlated potentials, the resulting phenomenology extends naturally across matter, acoustic, and electromagnetic waves, underscoring the prior discussion on the experimental accessibility of Berry crystals in several existing platforms. Furthermore, a Berry crystal can be regarded as an isotropic, multimode generalization of canonical one-dimensional Mathieu-type operators~\cite{Simon_Adv.Appl.Math_3_463_1982, Last_Commun.Math.Phys_164_421_1994, Jitomirskaya_Ann.Math_150_1159_1999}, which underlie a broad range of condensed-matter phenomena, including the Hofstadter butterfly~\cite{Hofstadter_Phys.Rev.B_14_2239_1976} and branched flow~\cite{Daza_PNAS_118_e2110285118_2021, Graf_Entropy_26_492_2024}. 

Our results further establish a connection between Berry-crystal localization and the physics of Bethe lattices, suggesting several directions for future work. These include extending the problem to three dimensions, where conventional Anderson systems exhibit mobility edges~\cite{Anderson_localization_book, Evers_Rev.Mod.Phys_80_1355_2008}, and exploring its relation to almost-periodic Schr{\"o}dinger operators and localization in one dimension~\cite{Simon_Adv.Appl.Math_3_463_1982}. On the other hand, it would be interesting to investigate whether slow temporal modulation of the potential can disrupt localization and give rise to Planckian diffusion, as proposed in Ref.~\cite{Broken_Anderson_localization}, and to determine how the extended critical states respond to such driving. More broadly, Berry crystals may offer an insightful framework for understanding wave transport in amorphous materials, where correlated forms of disorder arise naturally~\cite{Ziman_book}.


In summary, we have uncovered a distinct form of Anderson localization in a Berry crystal, interpolating between quasicrystalline order and uncorrelated disorder. Rather than exhibiting a sharp mobility edge, the system develops a broad critical regime separating the phases dominated by extended and localized states. This unusual mobility crossover can be understood through an effective description in terms of Bethe-lattice localization. Our results establish Berry crystals as a versatile platform for exploring how structured correlations in a potential can fundamentally reshape localization and mobility.

\section*{Acknowledgements}
The authors thank Xinchi Zhou, Zhongling Lu, Mingxuan Xiao, Yubo Zhang, Jasper Jain for discussions and suggestions. We thank the NSF Center for Integrated Quantum Materials (CIQM) Grant No. DMR-1231319, for support.

\section*{Appendix}
\subsection{Experimental scheme}\label{Appx:experimental_scheme} 

The Berry crystal can be realized for ultracold atoms by projecting a two-dimensional optical potential generated with a digital micromirror device~\cite{Gauthier2016}. The desired potential may also be generated by superposing $N$ mutually incoherent pairs of retroreflected plane waves. Incoherence can be achieved through small frequency detunings. Both $\beta$ and $q$ can be controlled. After preparing a Bose–Einstein condensed atomic cloud in the ground state of a harmonic trap, the Berry crystal potential can be ramped up slowly. A weak time-dependent perturbation can then produce a controlled energy excitation~\cite{Semeghinit_Nat.Phys.11_554_2015}. The subsequent expansion of the excited atoms following release of the trap provides a measure of the energy-dependent transport properties (Appx.~\ref{Appx:Wavepacket_analysis}). By varying the excitation frequency, one probes eigenstates at different energies of the Berry Hamiltonian. The energy dependence of the cloud size can be checked against our theory. The interactions can also be controlled using a magnetic Feshbach resonance~\cite{PhysRevLett.99.010403}. Effects of interactions on transport properties can be investigated, and the open problem of many-body localization can also be explored~\cite{RevModPhys.91.021001,nandkishore_many-body_2015}.

\subsection{Eigenstates and eigenvalues}\label{Appx:spectrum} 

We study the properties of the Berry crystal by first computing the eigenstates $\psi_n$ and eigenenergies $\varepsilon_n$ of its Hamiltonian 
\begin{equation}\label{Eq:Hamiltonian_appx}
    H = -\frac{\hbar^2}{2m}\nabla^2 + \frac{\beta}{\sqrt{N/2 \;}}\; \sum_{j=1}^N\cos(\mathbf{q}_j\cdot\mathbf{r}+\phi_j),
\end{equation} 
which is identical to Eq.~\eqref{Eq:Hamiltonian} in the main text. Here, $m$ denotes the effective electron mass, while the Berry potential (cf. Eq.~\eqref{eq:randompot}) is determined by the disorder strength $\beta$ and the number of random wave modes $N$. The wavevectors $\mathbf{q}_j = q(\hat{x}\cos\theta_j + \hat{y}\sin\theta_j)$ have a fixed magnitude $q$, whereas the wave directions $\theta_j$ and phases $\phi_j$ are independently drawn from a uniform distribution on $[0,2\pi]$. Throughout this work, we have chosen $q = 0.5\pi\,\textrm{Å}^{-1}$, and set the effective mass equal to the bare electron mass. Energies are expressed in units of $E_0 = \hbar^2 q^2 / (8m) \approx 2.35\,\textrm{eV}$, which naturally arises from the resonant-scattering condition discussed in the main text, and while length scales are measured in units of the characteristic Berry wavelength $\lambda = 2\pi/q = 4 \textrm{Å}$. Accordingly, both the potential strength $\beta$ and the system size $L$ are quoted in units of $E_0$ and $\lambda$, respectively. We emphasize, however, that the physical conclusions presented in this work are not tied to these particular parameter choices and remain valid after appropriate rescaling.  

The eigenvalue problem $H \psi_n = \varepsilon_n \psi_n$ is solved by utilizing an iterative Lanczos eigensolver with Dirichlet boundary conditions for a simulation box of size $L$. For each realization of the Berry potential, we compute all eigenstates within an energy window of width $\delta \varepsilon = 1.5\times10^{-2}$ centered at a target energy $E$. This window typically contains approximately $300$ eigenstates, providing sufficient statistical accuracy for the quantities considered throughout this work. The calculations are repeated for system sizes $L \in [200, 600]$, as required for the scaling analysis discussed in Sec.~\ref{Appx:scaling_analysis}. Unless stated otherwise, the level-statistics analysis is carried out for the representative system size of $L = 300$.

Finite systems inevitably support eigenstates that localize at the boundary of the simulation box, only disappearing in the thermodynamic limit. To prevent these spurious edge states from contaminating the statistical analysis, we remove them from the spectrum taking advantage of the binary-map filtering procedure introduced in Ref.~\cite{Zhu_Phys.Rev.A_109_013314_2024}. This filtering is particularly important for obtaining reliable spectral statistics, where even a small number of boundary states can noticeably distort the level-spacing distributions. It also improves the accuracy of the inverse participation ratio, although the corresponding analysis yielding the fractal dimension is found to be largely insensitive to the presence of these edge states. 

\subsection{Scaling analysis}\label{Appx:scaling_analysis} 

To characterize the emergence of a mobility crossover in a Berry crystal, we consider the generalized inverse participation ratio (IPR) of a normalized eigenstate $\psi_n(\mathbf{r})$, defined as
\begin{equation}\label{Eq:Appx_IPR_definition}
\Gamma_q^{(n)}=\int |\psi_n(\mathbf{r})|^{2q}\,d\mathbf{r}
\propto L^{-D_q(q-1)},
\end{equation}
where $D_q$ denotes the generalized fractal dimension~\cite{Girvin_book}. Throughout our work, we focus on the second moment ($q=2$), which is the standard quantity for distinguishing extended, critical, and localized states, see, e.g., Ref.~\cite{Fartash_Phys.Rev.B_113_195401_2026}.

For each target energy $E$, we evaluate $\Gamma_2^{(n)}$ for all eigenstates contained within a narrow energy window of width $\delta\varepsilon$, as described in Sec.~\ref{Appx:spectrum}, and subsequently introduce the corresponding ensemble averaged IPR, formally $\langle\Gamma\rangle=\langle\Gamma_2^{(n)}\rangle_n$. The fractal dimension $D_2(E)$ is then extracted from the finite-size scaling of the mean IPR employing system sizes $L\in[200,600]$. Specifically, we fit
\begin{equation}
\ln\langle\Gamma\rangle = D(0)-D_2\ln L,
\end{equation}
where $D(0)$ is a fitting constant. The resulting $D_2(E)$ is presented in Fig.~\ref{Fig:phase_diagram} for different Berry potential strengths $\beta$ at fixed mode number $N=60$. To quantify the overall localization tendency, we additionally define the mean fractal dimension $\bar{D}_2=\langle D_2(E)\rangle_{E\in[0.5,1.5]} $, which is shown as a function of both $\beta$ and $N$ in Fig.~\ref{Fig:phase_diagram_2}.

The extracted value of $D_2$ represents the average scaling behavior of eigenstates within the energy interval $\delta\varepsilon$. Although individual eigenstates exhibit slight variations in their individual fractal dimension, we find these fluctuations to be negligible, even in the vicinity of the mobility crossover. Moreover, repeating the analysis for higher moments ($q>2$) produces the same qualitative conclusion. We therefore conclude that the second-order IPR provides a reliable measure for identifying mobility crossovers in the Berry crystal.

\begin{figure}
    \centering
    \includegraphics[width=0.4\textwidth]{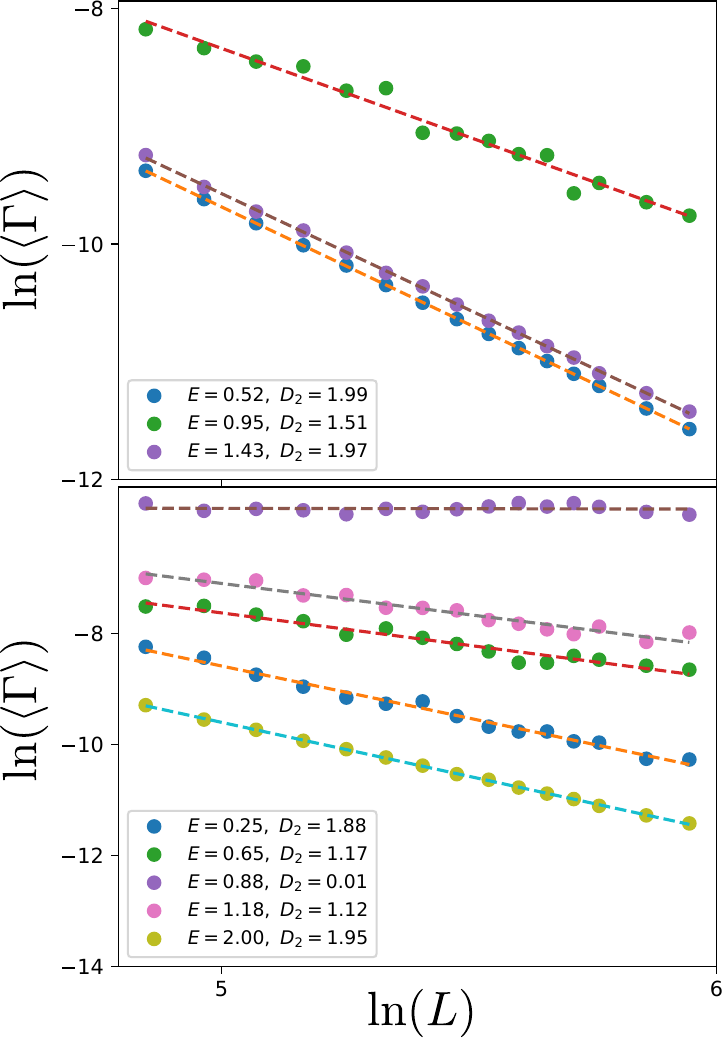}
    \caption{Finite-size scaling of the inverse participation ratio: Scaling of $\ln \langle\Gamma_2\rangle$ with system size $\ln L$ across different energy targets. The top panel covers a single-crossover regime at $\beta = 0.4$, illustrating a transition from extended states ($D_2 \approx 1.99$) through critical multifractal states near resonance ($D_2 \approx 1.51$) back to extended states ($D_2 \approx 1.97$). The bottom panel presents dual-crossover regime at $\beta = 0.9$, showcasing the full progression from extended states ($D_2 \approx 1.88$) to localized states near resonance ($D_2 \approx 0.01$), flanked by intermediate critical regimes.}
    \label{Fig:IPR_scaling}
\end{figure}

To further demonstrate this robustness of the extracted fractal dimensions, we explicitly present representative IPR scalings for the two distinct crossover scenarios observed in Fig.~\ref{Fig:phase_diagram}. In the single-crossover regime at $\beta=0.4$ [the upper panel of Fig.~\ref{Fig:IPR_scaling}], the scaling at $E \approx 0.5$ yields $D_2\simeq2$, characteristic of extended states. As the energy approaches the scattering resonance, the scaling becomes multifractal, with $D_2\approx1.5$ at $E \approx 0.95$, before reverting to an extended regime with $D_2\simeq2$ at $E \approx 1.45$. No localized states are observed throughout this crossover. By contrast, in the dual-crossover regime at $\beta=0.9$ [the lower panel of Fig.~\ref{Fig:IPR_scaling}], the low-energy states are initially extended ($D_2\simeq2$ at $E \approx 0.25$), evolve through a critical regime, and become localized near the scattering resonance, where $D_2\simeq0$. At higher energies, for example at $E \approx 1.2$, the scaling again becomes multifractal, signaling a re-emergence of critical states. These representative examples thus corroborate the phase diagram of Fig.~\ref{Fig:phase_diagram} obtained from the finite-size scaling analysis.

\subsection{Level statistics} \label{Appx:level_statistics} 

The simplest statistical indicator of quantum chaos is the nearest-neighbor level-spacing distribution, which characterizes the distribution of spacings between consecutive (unfolded) energy levels and therefore probes short-range spectral correlations. In this work, however, we employ the closely related gap-ratio statistics. Given the ordered eigenenergies ${\varepsilon_n}$ of the Hamiltonian in Eq.~\eqref{Eq:Hamiltonian_appx}, we define the ratio of consecutive spacings
\begin{equation}
    \eta_n =\frac{\varepsilon_{n + 1} - \varepsilon_n}{\varepsilon_{n} - \varepsilon_{n- 1}}
\end{equation} 
A principal advantage of the gap-ratio approach is that it is independent of the local density of states, and therefore does not require spectra unfolding, which can introduce subtle systematic errors into numerical analyses (see, e.g., Ref.~\cite{Gomez_Phys.Rev.E_66_036209_2022}). Consequently, the gap ratio has become one of the standard diagnostics of quantum chaos, particularly for systems with an energy-dependent density of states.

Within the context of quantum chaos, spectral statistics are commonly interpreted through the BGS conjecture~\cite{Bohigas_phys.rev.lett_52_1_1984}, which relates quantum spectra to the nature of the underlying classical dynamics (see, e.g., Refs.~\cite{Heller_book, Stockmann_book, Tabor_book, Haake_book, Gutzwiller_book}): Integrable systems possess essentially uncorrelated energy levels and thus exhibit Poisson statistics, whereas classically chaotic systems are described by random-matrix theory. Although exact expressions exist for the Gaussian random-matrix ensembles~\cite{Mehta_book}, such as the Gaussian orthogonal ensemble (GOE) appropriate for time-reversal-symmetric systems, we employ the widely employed Wigner-surmise approximation.

For the gap-ratio distribution, the Poisson and GOE limits are given by
\begin{equation*}
    P_{\textrm{S}}(\eta) = \frac{1}{(1 + \eta)^2} 
    \quad \textrm{and} \quad
    P_{\textrm{GOE}} = \frac{27 \eta (1 + \eta)}{8(1 + \eta + \eta^2)^{5/2}},
\end{equation*}
respectively. These distributions exhibit qualitatively different asymptotic behavior. For large ratios, $\eta \gg 1$, we have $P_{\textrm{S}}(\eta) \sim \eta^{-2}$ and $P_{\textrm{GOE}} \sim \eta^{-3}$. On the other hand, for small values of $\eta \ll 1$ the GOE statistics exhibit level repulsion, i.e, suppressing small spacing ratio, in contrast to the Poisson case, formally $P_{\textrm{S}}(\eta) \rightarrow 1$ and $P_{\textrm{GOE}} \rightarrow 0$ when $\eta \rightarrow 0$. The suppression of small spacing ratios therefore provides a direct signature of chaotic spectral correlations.

For numerical stability, it is often convenient to focus on the symmetrized ratio $\tilde{\eta} = \min \{\eta_n, 1/\eta_n\}$, and subsequently compute the mean symmetrized ratio $\langle \tilde{\eta}_n \rangle$ over the studied spectrum $\{ \varepsilon_n \}$. The mean values for the Poisson and GOE limits are analytically known to be $\tilde{\eta}_{\mathrm{P}} \approx 0.386$ and $\tilde{\eta}_{\mathrm{GOE}} \approx 0.536$, respectively. Following the essence of the BGS conjecture, we subsequently define a normalized ``chaoticity'' indicator $\chi$ as follows 
\begin{equation}\label{Eq:chaos_estimators_1}
    \chi = \frac{\langle \tilde{\eta}_n \rangle - \tilde{\eta}_{\textrm{P}}}{\tilde{\eta}_{\textrm{GOE}} - \tilde{\eta}_{\textrm{P}}}
\end{equation}
such that $\chi = 0$ and $\chi = 1$ correspond to the Poissonian (regular) and GOE (chaotic) limits, respectively. Figure~\ref{Fig:level_statistics} shows the resulting chaos estimator $\chi$  as a function of the energy $E$ and the Berry-potential strength $\beta$. 

In complement to the gap-ratio analysis above, we also characterize spectral correlations employing the variance of the unfolded level spacings, defined as
\begin{equation}
    \mu = \left \langle \left( \frac{\varepsilon_{n + 1} - \varepsilon_n }{g(E)} \right)^2 \right \rangle - \left \langle \left( \frac{\varepsilon_{n + 1} - \varepsilon_n }{g(E)} \right) \right \rangle^2,
\end{equation}
where $g(E)$ is the local density of states in the vicinity of the studied energy $E$ and the averaging is again carried out over the studied spectrum $\{ \varepsilon_n \}$. For Poisson and GOE ensembles, the corresponding analytical values of spectral variance are $\mu_{\textrm{P}} = 1$ and $\mu_{\textrm{GOE}} =4/\pi-1$, respectively.

Unlike the gap ratio, the spectral variance measure depends on the unfolding of the spectrum via the density of states. Since our analysis is restricted to a relatively narrow energy window $\delta\varepsilon$, numerical calculations demonstrate that the density of states varies only weakly within this interval. We therefore approximate it by its local average
\begin{equation*}
    g(E) \approx \frac{\#\{\varepsilon_n \, \vert \, \vert \varepsilon_n - E \vert \le \delta \varepsilon \}}{\delta \varepsilon}. 
\end{equation*}
As described in Sec.~\ref{Appx:spectrum}, approximately $300$ eigenstates are computed within an energy window of width $\delta\varepsilon\sim10^{-2}$ centered on each target energy $E$, providing the local estimate of the density of states through the approximation introduced above.

\begin{figure}[h!]
    \centering
    \includegraphics[width=0.5\textwidth]{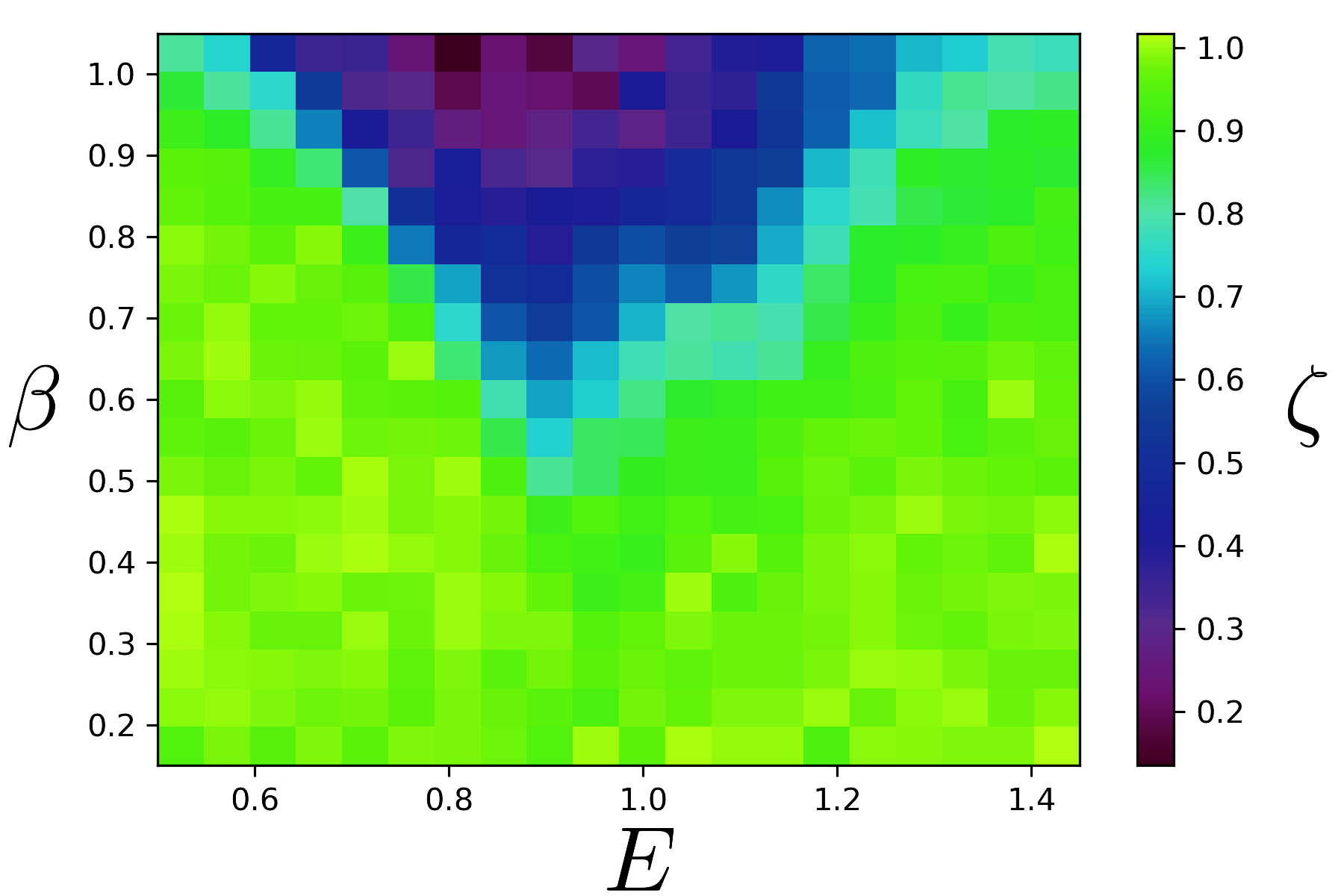}
    \caption{Spectral variance chaos indicator: Chaos indicator $\zeta(E, \beta)$ derived from the level-spacing variance $\mu$ [see Eq.~\eqref{Eq:chaos_estimators_2}] as a function of energy $E$ and Berry potential strength $\beta$. Values of $\zeta \approx 1$ denote Wigner-Dyson (GOE) statistics characteristic of ergodic states, whereas $\zeta \approx 0$ corresponds to Poisson statistics indicative of localized states. The spectral variance map exhibits strong agreement with the gap-ratio analysis [see Fig.~\ref{Fig:phase_diagram}], indicating only the conventional, Anderson-like transition instead of the dual crossover observed in the inverse participation analysis of eigenstates.}
    \label{Fig:spectral_variance}
\end{figure}

In complete analogy with Eq.~\eqref{Eq:chaos_estimators_1}, we introduce another chaos indicator based on the spectral variance as
\begin{equation}\label{Eq:chaos_estimators_2}
    \zeta = \frac{\mu - \mu_{\textrm{P}}}{\mu_{\textrm{GOE}} - \mu_{\textrm{P}}},
\end{equation} 
so that $\zeta=0$ stands for Poisson statistics and $\zeta=1$ for the GOE limit. The resulting map of $\zeta(E,\beta)$, shown in Fig.~\ref{Fig:spectral_variance}, is in excellent agreement with the gap-ratio analysis, illustrated in Fig.~\ref{Fig:level_statistics}. Similarly to the gap-ratio analysis, Poisson statistics of the indicator $\zeta$ signals localized eigenstates, whereas GOE statistics is characteristic of ergodic extended states. However, we find that neither $\chi$ nor $\zeta$ reliably detects the critical states present in the Berry crystal. 

Consequently, although both $\chi$ and $\zeta$ accurately identify the onset of Anderson localization, they fail to resolve the broad intermediate regime of critical states revealed by the inverse-participation-ratio analysis (see, e.g., Sec.~\ref{Appx:scaling_analysis}). Instead, the spectral statistics suggests a sharper mobility crossover separating localized and extended domains than observed in the IPR analysis, thereby obscuring the existence of the intervening critical regime. This limitation is particularly evident near the resonant energy $E\approx1$ for weak Berry potentials ($\beta \lesssim 0.4$), where the system evolves continuously between two extended-state phases through a broad region of multifractal critical states without exhibiting Anderson localization.

\subsection{Wavepacket analysis}\label{Appx:Wavepacket_analysis}

To complement the static IPR analysis discussed in Sec.~\ref{Appx:scaling_analysis}, we investigate the localization properties of the Berry crystal from the perspective of wavepacket dynamics. To this end, we consider the time evolution of spectrally narrow Gaussian wavepackets of the form
\begin{equation}\label{Eq:wavepacket_appx}
\psi(\mathbf r,t = 0) = \mathcal{N}
\exp\left(
-\frac{|\mathbf r-\mathbf r_0|^2}{2\sigma^2}
+i\mathbf k_0\cdot\mathbf r
\right),
\end{equation}
where $\mathcal N$ denotes the normalization constant. Throughout this analysis, we employ a square simulation cell of size $L=320\lambda$, with $\lambda = 4 \textrm{Å}$, consistent with the main text. The wavepacket is centered at $\mathbf r_0 = (0,0)$ and has spatial range of $\sigma = 19\lambda$, corresponding to a spectral width spanning several hundred eigenstates around the target energy $E$ (measured in units of $E_0 \approx 2.35 \,\mathrm{eV}$). The initial momentum is chosen according to $|\mathbf k_0|=\sqrt{2mE}/\hbar$, ensuring that the mean energy of the wavepacket coincides with the target energy.

The time evolution is generated by the Berry-crystal Hamiltonian [see Eq.~\eqref{Eq:Hamiltonian_appx}] employing the split-operator method (see, e.g., Refs.~\cite{Heller_book,aydin_proc.natl.acad.sci_121_e2404853121_2024, Graf_Entropy_26_492_2024,zimmermann_entropy_26_552_2024}) with periodic boundary conditions. Unless stated otherwise, we consider a Berry potential with $\beta=0.4$, and $N=60$, thereby matching with the IPR analysis presented in Fig.~\ref{Eq:Hamiltonian_appx}. The propagation extends to a final time of $T=4000t_0$, where $t_0=\hbar/E_0 \approx 0.280\,\mathrm{fs}$, corresponding to $3\times10^4$ time steps with temporal resolution $0.037\,\mathrm{fs}$. 

To characterize the dynamics, we compute the mean-square displacement (MSD), defined as
\begin{equation*}
    \langle r^2(t)\rangle = \int |\mathbf r-\mathbf r_0|^2
|\psi(\mathbf r,t)|^2 \, \textrm{d} \textbf{r},
\end{equation*}
together with the instantaneous (diffusion) exponent 
\begin{equation*}
    \alpha(t) = \frac{\textrm{d}\ln \langle r^2(t)\rangle}{\textrm{d}\ln t}.
\end{equation*}
For asymptotically long times, this quantity establishes a classification of diffusive transport, and we therefore adopt the associated nomenclature for the instantaneous MSD exponent: $\alpha\simeq2$ corresponds to ballistic motion, $\alpha=1$ to normal diffusion, $1<\alpha<2$ to superdiffusion, $0<\alpha<1$ to subdiffusion, and $\alpha\simeq0$ to the absence of transport. In the present context, superdiffusive behavior reflects the dominance of extended eigenstates within the wavepacket, whereas subdiffusive dynamics indicate an increasing contribution from critical or localized states.

\begin{figure}
    \centering
    \includegraphics[width=0.5\textwidth]{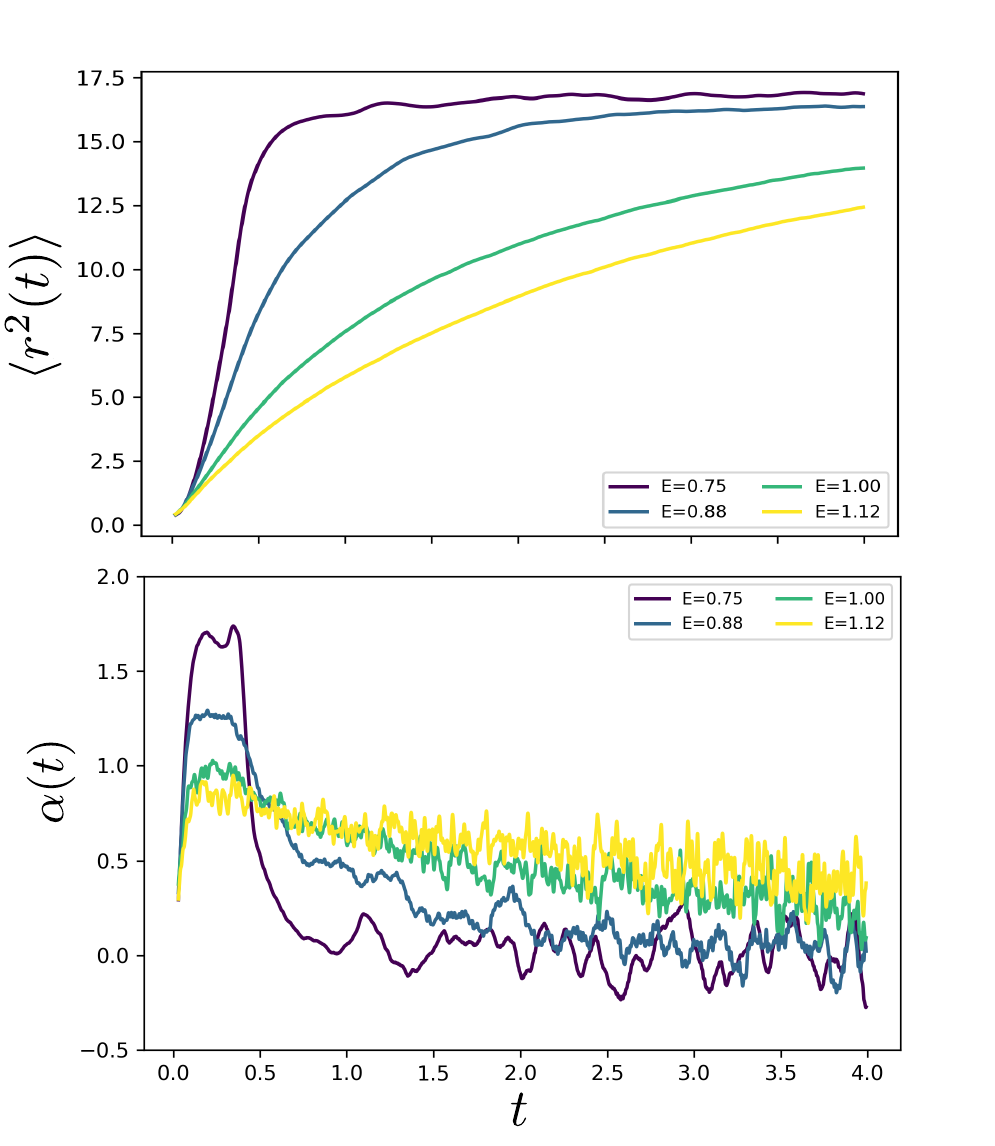}
    \caption{Wavepacket dynamics across the mobility crossover: Mean-square displacement $\langle r^2(t) \rangle$ (top panel) and instantaneous diffusion exponent $\alpha(t)$ (bottom panel) for spectrally narrow Gaussian wavepackets at $\beta = 0.4$ centered at target energies $E \in \{0.75, 0.88, 1.00, 1.12\}$. Lower-energy wavepackets ($E=0.75$) display initial superdiffusive transport ($\alpha > 1$) before boundary saturation, whereas wavepackets centered within the critical regime ($E=1.00, 1.12$) exhibit persistent subdiffusive dynamics ($\alpha < 1$).}
    \label{Fig:wavepacket_diffusion}
\end{figure}

Figure~\ref{Fig:wavepacket_diffusion} presents the MSD and the corresponding diffusion exponent for wavepackets centered at four representative target energies $E \in \{0.75, 0.88, 1.00, 1.12\}$, selected to probe the mobility crossover identified by the IPR analysis at $\beta=0.4$ (see Fig.~\ref{Fig:phase_diagram}). During the initial stage of the evolution ($t\lesssim10^3t_0$), all wavepackets spread rapidly, leading to large values of the MSD and an approximately constant diffusion exponent. This short-time behavior is governed by the most extended eigenstates contained in the spectral composition of the wavepacket. Consequently, the wavepackets with $E=0.75$ and $0.88$ initially exhibit superdiffusive transport, whereas those centered at $E=1.00$ and $1.12$ are already subdiffusive. As these rapidly propagating components traverse the finite simulation cell, they eventually explore the entire periodic domain, causing their contribution to the MSD to saturate. This saturation is then accompanied by oscillations in the diffusion exponent $\alpha$, originating from the interference of the wavepacket returning through the periodic boundaries. 

At later times, the wavepacket dynamics become increasingly dominated by slower spectral components, resulting in a gradual reduction of the effective diffusion exponent. The wavepacket with $E=0.75$ provides the clearest example. After reaching the system boundaries at approximately $t \sim 800$, its diffusion exponent rapidly decreases toward zero, indicating that the initial spreading was almost entirely governed by highly extended states. Once these components have explored the entire simulation cell, only weak residual spreading remains, producing a slow decay of $\alpha(t)$ joined by the self-interference oscillations due to periodic boundary condition. This behavior is fully consistent with the IPR analysis, which identifies the states $E \lesssim 0.85$ as extended.

The remaining three wavepackets, on the other hand, exhibit qualitatively different dynamics. Since their spectral composition contains an increasing fraction of critical and partially localized eigenstates, the MSD saturates on a substantially longer time scale. The wavepacket centered at $E=0.88$, located in the vicinity of the mobility crossover, initially remains superdiffusive, although less strongly than the wavepacket with $E=0.75$. After $t \sim 700$, its diffusion exponent, however, drops to $\alpha \sim 0.5$, before gradually approaching zero at longer times. By contrast, the wavepackets geared for $E=1.00$ and $1.12$, which both lie well inside the critical regime identified by the IPR analysis, remain close to normal diffusion during the initial stage before continuously evolving into increasingly subdiffusive behavior. Unlike the lower-energy wavepackets, they exhibit little evidence of the long-time oscillations associated with rapidly returning extended components, indicating that their evolution is dominated by slowly propagating eigenstates with significant spatial localization.

Overall, the analysis on wavepacket dynamics provides an independent dynamical characterization of the mobility crossover inferred from the finite-size scaling of the IPR or the level statistics, discussed in Sec.~\ref{Appx:scaling_analysis} and Sec.~\ref{Appx:level_statistics}, respectively. The gradual evolution from superdiffusive to subdiffusive transport closely follows the changes in the spectral composition of the wavepackets, thereby supporting the phase diagram obtained from the static fractal-dimension analysis. At the same time, the dynamical simulations reveal the transport signatures associated with the different mobility regimes and illustrate their consequences for the long-time propagation of mobile carriers in the Berry crystal. 

\subsection{Perturbation series approach}\label{Appx:Perturbation_theory}

Perturbation theory provides a simple framework for identifying the transition between extended and localized eigenstates: localization in coordinate space necessarily translates into delocalization in momentum space, and vice versa. Consequently, a state at energy $E$ is expected to be spatially localized whenever the Born expansion about the energy shell $\vert p \vert \approx \sqrt{2m E}$ fails to converge.

To make this argument quantitative, we treat the Berry potential as a weak perturbation. The unperturbed basis consists of plane waves $|\mathbf{k}\rangle$, which we employ to construct the exact eigenstate $|\Psi\rangle$ of the Hamiltonian $H$ [see Eq.~\eqref{Eq:Hamiltonian}]. Based on the Lippmann-Schwinger equation, we can write
\begin{equation}
|\Psi\rangle = \sum_{m=0}^{\infty} (\hat{G}_0(E)\hat{V})^m |\mathbf{k}\rangle,
\label{eq:born_series}
\end{equation}
where the free propagator of the system is 
\begin{equation}
    G(E) = \left(E - \frac{p^2}{2m} + i\eta \right)^{-1}
\end{equation}
with an infinitesimal positive parameter $\eta >0$, and $\hat{V}$ denotes the operator associated with the Berry potential, i.e., $V(\mathbf{r}) = \langle \mathbf{r} \vert \hat{V} \vert \mathbf{r \rangle}$ [see Eq.~\eqref{eq:randompot}]. If the Born series above converges, the exact eigenstate remains continuously connected to the unperturbed plane wave, establishing the existence of an extended state at energy $E$. Conversely, divergence of the series marks the onset of localization through the breakdown of perturbation theory. We determine this (mobility) transition between localized and extended states numerically by analyzing the convergence of the series in Eq.~\eqref{eq:born_series}, yielding the phase boundary shown as the blue curve in Fig.~\ref{Fig:phase_diagram}.

More specifically, we consider a finite system of size $L=100$, and choose the energy broadening to coincide with the finite-size level spacing on the resonant momentum shell, i.e., $\eta=(\hbar^2q/2m)(2\pi/L)$. Starting from the normalized plane-wave state $|\psi^{(0)}\rangle=|\mathbf{k}\rangle$ with wave vector $\vert \boldsymbol{k} \vert =\sqrt{2mE/ \hbar^2}$, we iteratively generate the Born series according to 
\begin{equation}
    |\psi^{(m+1)}\rangle= G_0(E) \hat{V} |\psi^{(m)}\rangle.
\end{equation}
where $G_0(E)$ is the free propagator and $\hat{V}$ denotes the Berry potential. To quantify the convergence of the expansion, we introduce an asymptotic growth factor
\begin{equation}
    r_m=\frac{\|\psi^{(m+1)}\|}{\|\psi^{(m)}\|},
\end{equation}
where the norm is induced by the standard inner product of the Hilbert space. Numerically, we find that after a sufficient number of iterations, the asymptotic growth factor $r_m$ seems to approach a well-defined value $r(E,\beta)$ that depends only on the energy $E$ and the disorder strength $\beta$ (we here use $m=100$; the numerical results have been verified to be converged by repeating the calculations with $m = 150$ and $m = 200$). 

When this estimation satisfies $r < 1$, the successive terms of the Born series decrease in norm, indicating convergence of the perturbative expansion about the free plane-wave state. Conversely, a result $r > 1$ signals an instability of the expansion. Even though the divergence of the Born series alone does not constitute a rigorous proof of localization, its convergence provides a sufficient condition for the existence of extended states, and serves as a practical estimate for the boundary of the extended regime, beyond which the perturbative description in terms of plane waves ceases to be valid. The contour defined by $r(E,\beta) = 1$ is presented in Fig.~\ref{Fig:phase_diagram}(a).

\subsection{Bethe lattice approach}\label{Appx:Bethe_lattice} 

While the convergence analysis of the Born series discussed in Sec.~\ref{Appx:Perturbation_theory} provides a lower-bound estimate for the boundary of the localization region shown in Fig.~\ref{Fig:phase_diagram}, a sharper estimate of the mobility transition can be obtained within an effective Bethe-lattice description. In a nutshell, we map the Berry crystal onto a Cayley tree by retaining the dominant scattering processes in momentum space. The resulting graph can then be interpreted as an effective Bethe lattice, yielding a tight-binding model of the Anderson type in which the disorder strength is determined by the energy detuning associated with the Berry potential. 

To obtain a more accurate estimate than that provided by the Born-series analysis, we express the Berry-crystal Hamiltonian $H$ of Eq.~\eqref{Eq:Hamiltonian_appx} in the momentum basis $|\mathbf k \rangle$, resulting in
\begin{equation}
\begin{aligned}
    \langle \mathbf k | H | \mathbf k' \rangle = \delta_{\mathbf k,\mathbf k'} \epsilon_\mathbf k + V_{\mathbf k,\mathbf k'}
\end{aligned}
\label{eq:H_k}
\end{equation}
where the kinetic energy is $\varepsilon_{\mathbf{k}} = \hbar^2 \vert \mathbf{k} \vert^2/2m$ and the potential energy related to the Berry potential is
\begin{equation*}
V_{\mathbf k,\mathbf k'} = \frac{\beta}{2} \sum_{j = 1}^N \left[ e^{i\phi_j}\delta_{\mathbf k',\mathbf k+\mathbf q_j}
+ e^{-i\phi_j}\delta_{\mathbf k',\mathbf k-\mathbf q_j} \right],
\end{equation*}
where $\beta$ and $\phi_j$ are the strength of the potential and the (random) phases of the modes in the potential, respectively. We now see that the Berry potential only scatters by vectors $\pm\mathbf q_j$. Conversely, we can re-interpret this situation as a tight-binding model on a highly connected graph in momentum space. 

\begin{figure}[h!]
    \centering
    \includegraphics[width=0.45\textwidth]{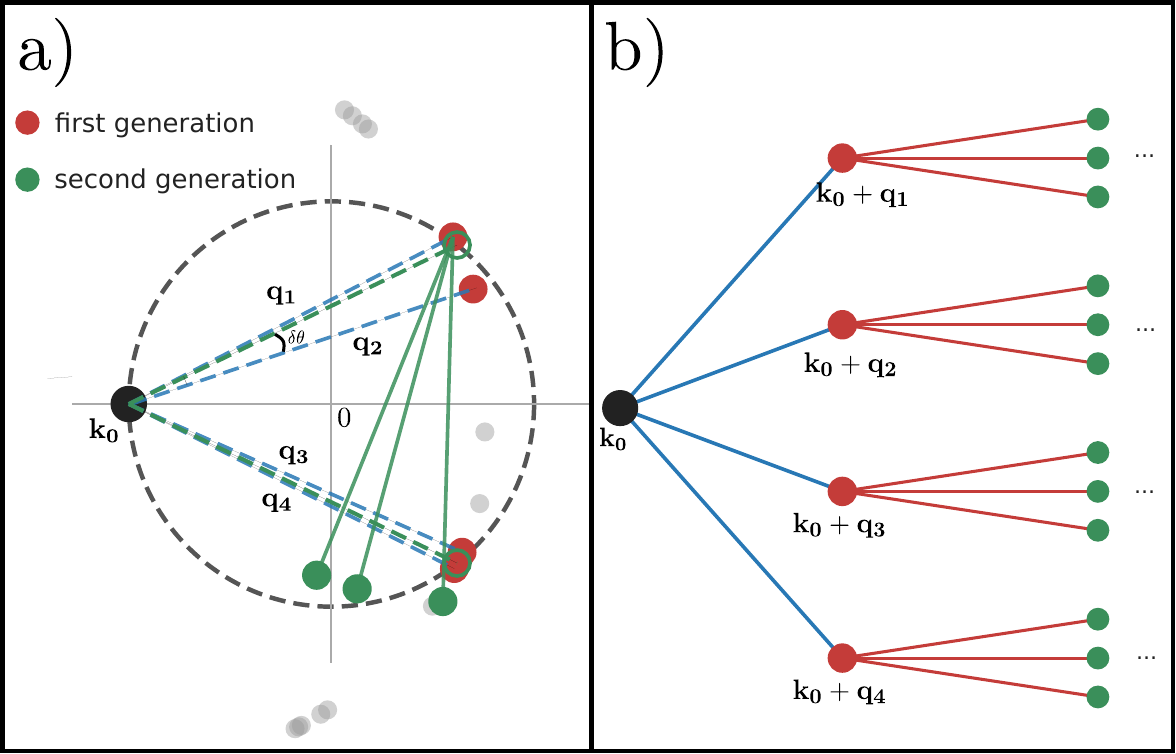}
    \caption{Mapping Berry-crystal scattering onto an effective Bethe lattice: (a) Momentum-space scattering on the resonant shell \(|\mathbf{k}|=k_0\) (dashed circle). The initial state \(\mathbf{k}_0\) (black) scatters via \(\mathbf{q}_{1} \dots \mathbf{q}_{4}\) into near-resonant first-generation states \(\mathbf{k}_0+\mathbf{q}_j\) (red), detuned by \(\delta\theta\) [see Eq.~\eqref{eq:bragg}]. Subsequent scattering produces second-generation states (green), yielding locally tree-like connectivity; gray points denote non-resonant states. (b) Effective graph, with \(\mathbf{k}_0\) as the root of a Cayley tree with branching number \(K\simeq3\). The tight-binding Hamiltonian \(H_{\mathrm{eff}}\) [see Eq.~\eqref{Eq:TB_Hamiltonian_appx}] realizes an Anderson model on a Bethe lattice, with generation-dependent onsite disorder set by the angular detuning.}\label{Fig:Bethe_lattice_schematic}
\end{figure}

When expanding a generic state at a given target energy $E$ as $\vert \Psi \rangle = \sum_{\mathbf{k}} c_{\mathbf{k}} \vert \mathbf{k}\rangle$, the coefficients $c_{\mathbf{k}}$ are determined by the following Schrödinger equation
\begin{equation}
    (E - \varepsilon_{\mathbf k}) c_\mathbf k = \sum_{\mathbf k'} V_{\mathbf{k}\mathbf{k}'} c_{\mathbf{k}'}
\end{equation}
which demonstrates that the components with momenta $\mathbf k$ with $|\varepsilon_{\mathbf k}-E|\lesssim |\beta|$ carry the most of the weight. Therefore, we can focus on a momentum shell defined by the energy $E$; the components outside this shell are largely suppressed. Further, with a fixed energy \(E_{k_0}=E(k_0)\), the Hamiltonian can be projected onto a thin annulus in momentum space, as illustrated in Fig.~\ref{Fig:Bethe_lattice_schematic}(a). Within this projection, it acts primarily on momentum states lying close to the circle \(|k|=k_0\). The diagonal terms of the Hamiltonian correspond on-site energies $\varepsilon_{\mathbf k}$, while off-diagonal terms with constant amplitudes correspond to hoppings between sites whose momentum difference equals one of the $\mathbf q_j$. In other words, the system dynamics is mapped onto an effective graph. 

The key quantity controlling the structure of the effective graph is the energy mismatch $\Delta E(\mathbf k,\mathbf q)$ associated with the scattering process of $\mathbf k\to\mathbf k+\mathbf q$, which is
\begin{equation}
\Delta E(\mathbf k,\mathbf q)
=
\frac{|\mathbf k+\mathbf q|^2 - k_0^2}{2m}
=
\frac{q^2}{2m} + \frac{k_0 q}{m}\cos\theta,
\label{eq:bragg}
\end{equation}
where $\theta$ is the angle between $\mathbf k$ and $\mathbf q$. For $k_0>q/2$, the equation $\cos\theta_0 = -\frac{q}{2k_0}$ has two real solutions which give two Bragg angles. Since the potential has only a finite number of peaks in momentum space, the scattering is never perfectly resonant. For each Bragg angle, there are two nearly resonant scattering processes. Therefore, each momentum $\mathbf k$ is connected to four neighbors by nearly resonant scattering processes. Repeated scattering generates a Cayley tree structure because loops are prevented by the incommensurate $\mathbf q_j$, and each momentum, i.e, each graph point, have the same local connectivity. In essence, the effective graph associated with our Berry potential is a Cayley tree with branching number $K\simeq 3$, as represented schematically in Fig.~\ref{Fig:Bethe_lattice_schematic}(b).

We can now interpret this momentum-space graph as an effective Bethe lattice and formulate the corresponding tight-binding model. Denoting the momentum states by $\vert i \rangle$ ($i = 1, \dots, N$), we write
\begin{equation}\label{Eq:TB_Hamiltonian_appx}
H_{\rm{eff}}
=
\sum_{i = 1}^N \Delta_{i}\,|i\rangle\langle i|
+
\sum_{\langle i,j\rangle}
t\,e^{i\phi_{ij}} |i\rangle\langle j|
+ \text{h.c.}, 
\end{equation}
where $t = \beta/\sqrt{2N}$ is the hopping amplitude, $\phi_{ij}$ are random phases, and $\Delta_i=\varepsilon_i- E_{k_0}$ denotes the on-site energy measured relative to the energy of the momentum shell. The on-site energies are determined by the angular detuning $\delta\theta$ from the Bragg angle $\theta_0$. For small $\delta\theta$, we obtain 
\begin{equation}
\varepsilon_{i}
\sim E_{k_0} + 2 k_0 q\sin\theta_0\,\delta\theta
\sim E_{k_0} + 8 E_0 \sqrt{E_{k_0}/E_0-1}\,\delta\theta, 
\end{equation}
where the kinetic energy associated with the momentum shell is $E_{k_0}=\hbar^2 k_0^2/2m$. Consequently, in the resonant regime, $k_0>q/2$, the effective on-site disorder is linear in the angular detuning, meaning $\Delta_i \sim 8E_0 \sqrt{E_{k_0}/E_0-1}\delta\theta$.  

The effective description however changes qualitatively when the Bragg resonance ceases to exist. For $k_0<q/2$, Eq.~\eqref{eq:bragg} has no real solution, and all first-order scattering processes are therefore off shell. The leading contribution to the effective Hamiltonian consequently arises from second-order virtual scattering. These processes connect each momentum state to four neighboring states and, importantly, preserve the same tight-binding structure as Eq.~\eqref{Eq:TB_Hamiltonian_appx}. The main difference is that both the hopping amplitude and the on-site disorder are renormalized, given by $t \sim \frac{\beta^2 E_0} {8N\left(1-\sqrt{E_{k_0}/E_0}\right)}$ and $\Delta_i \sim 4E_0 \left( 2 -\sqrt{E_{k_0}/E_0} \right) (\delta\theta)^2$, respectively.  

The resonant and nonresonant regimes can therefore be described within a unified Bethe-lattice framework of the form given by Eq.~\eqref{Eq:TB_Hamiltonian_appx}, with the effective disorder arising from first-order angular detuning in the resonant regime and from second-order virtual processes in the nonresonant regime. It is however convenient to normalize the effective Hamiltonian by the corresponding hopping amplitude and express all energies in units of $E_0$, yielding
\begin{equation}\label{Eq:TB_Hamiltonian_2_appx}
H_{\rm{eff}}
=
\sum_{i = 1}^N \tilde{\Delta}_{i}\,|i\rangle\langle i|
+
\sum_{\langle i,j\rangle}
\,e^{i\phi_{ij}} |i\rangle\langle j|
+ \text{h.c.}, 
\end{equation} 
where the dimensionless onsite energies $\tilde{\Delta}_{i}$ scales as
\begin{equation}
\tilde{\Delta}_{i} \sim  
\begin{cases}
16 \left( 2 - \sqrt{E} \right) \left( 1 - \sqrt{E} \right) N (\delta \theta)^2, & E > 1, \\
\\
8  \sqrt{E - 1} \delta \theta, & E < 1.
\end{cases}
\end{equation}
The typical angular detuning is set by the spacing between neighboring scattering directions, $\delta\theta\sim\frac{2\pi}{mN}$, where $m$ is the number of connecting neighbors in the effective lattice.

In practice, we treat the on-site energies $\tilde{\Delta}_i$ as random variables whose characteristic scale is determined by the above estimates. We then generate their distribution numerically from $5000$ independent realizations of the Berry potential [see, e.g., Eq.~\eqref{Eq:Hamiltonian_appx}]. The resulting distribution is well approximated by
\begin{equation}\label{Eq:onsite_distribution_appx}
P(\tilde{\Delta}) \approx
\begin{cases}
\frac{1}{W}\exp(-\frac{\tilde{\Delta}}{W}), & E > 1, \\
\\
\frac{1}{\sqrt{2\pi}W} \exp(-\frac{\tilde{\Delta}^2}{2W^2}), & E < 1,
\end{cases}
\end{equation} 
with the following standard deviation
\begin{equation}\label{Eq:onsite_distribution_width_appx} 
W(E, \beta) \approx
\begin{cases}
\dfrac{46\pi^2}{N \beta^2} (2 - \sqrt{E})(1 - \sqrt{E}), & E > 1, \\
\\
\dfrac{16\pi}{5\beta}\sqrt{\dfrac{2(E - 1)}{N}}, & E < 1.
\end{cases}
\end{equation}
This effective model hence coincides with an Anderson model on a Bethe lattice with connectivity  $K=3$, which is known to exhibit a mobility transition as the disorder strength is varied~\cite{Abou_J.Phys.C_6_1734_1973, Monthus_J.Phys.A_44_145001_2011}.

In the present problem, the relevant disorder strength is set by the variance $W(E,\beta)$, which varies systematically with target energy $E$ and Berry potential strength $\beta$ for a given mode number $N$. In particular, states far from the resonant energy, $E \gg 1$, experience a large effective disorder and are consequently localized on the momentum-space graph. Momentum-space localization on the other hand corresponds to spatially extended states in coordinate space. Conversely, as the resonant energy is approached, $W(E,\beta)$ decreases and momentum-space states become increasingly extended. This regime can support non-ergodic extended states, while sufficiently weak disorder ultimately leads to real-space localization.

We characterize the mobility crossover by a critical disorder strength $W_C$, such that $W \geq W_C$ agree to localization on the momentum-space graph and, consequently, to extended states in coordinate space. For the connectivity $K=3$ relevant to our model, we take $W_C=6$, following the value reported for the Cayley/Bethe lattice~\cite{Abou_J.Phys.C_6_1734_1973, Monthus_J.Phys.A_44_145001_2011}. As illustrated by the red dashed curve in Fig.~\ref{Fig:phase_diagram}, the contour $W(E,\beta) = 6$ provides a substantially sharper estimate of the mobility transition than the criterion based on the convergence of the Born series discussed in Sec.~\ref{Appx:Perturbation_theory}. 

The Bethe-lattice description further predicts a representative scaling of the mobility crossover with the number of modes $N$. Based on Eq.~\eqref{Eq:onsite_distribution_width_appx}, we conclude $\beta\propto\frac{1}{\sqrt{N}}$ for the mobility crossover line. This behavior agrees well with the scaling of the mean fractal dimension $\bar{D}_2(\beta,N)$ shown in Fig.~\ref{Fig:phase_diagram_2}. The agreement, nevertheless, is limited to approximately $N \lesssim 100$, beyond which systematic deviations become apparent. This divergence is expected since the effective Hamiltonian in Eq.~\eqref{Eq:TB_Hamiltonian_2_appx} entails only first- and second-order scattering processes. As $N$ increases, the detuning decreases as $\delta\theta\sim1/N$, causing higher-order virtual processes to become increasingly important. Hence, in the limit $N\rightarrow\infty$, a complete description of the emergence of a mobility crossover requires the inclusion of scattering processes beyond second order, which are not considered in this work.

\bibliography{references}

\end{document}